\documentstyle[aps,psfig,epsfig]{revtex}
\newcommand{\an}[0]{\bigtriangledown}
\begin{document}

\draft
\begin{center}
{\Large {\bf \sc the radiative decay of vector mesons}}
\\[5mm]
Ting-Liang Zhuang\\
{\small Center for Nonlinear Science, University of Science and Technology
of China}\\
{\small Hefei, Anhui 230026, People's Republic of China}\\[2mm]
Xiao-Jun Wang\\
{\small Center for Fundamental Physics, University of Science and
Technology  
of China}\\
{\small Hefei, Anhui 230026, People's Republic of China}\\[2mm]
Mu-Lin Yan\\
{\small Center for Advanced Study \\
Tsinghua University, Beijing, 100084, People's Republic of China}\\
{\small and}\\
{\small  Center for Fundamental Physics,
University of Science and Technology of China\\
Hefei, Anhui 230026, People's Republic of China}\footnote{mailing
address}\\
\end{center}
\vspace{8mm}
\begin{abstract}
\noindent
In this paper, radiative decays $\rho^0 \rightarrow \pi^+\pi^-\gamma ,
\pi^0\pi^0\gamma$ ,$\phi \rightarrow K^+K^-\gamma ,K^0 \bar{K^0}\gamma$
are studied systematically in the U(3)$_L\times$U(3)$_R$ chiral 
theory of mesons. The theoretical differential spectrum with respect 
to photon energy and branch ratio for $\rho^0 \rightarrow
\pi^+\pi^-\gamma$ agree well with the experimental data. Differential
spectrums and branch ratios for $\rho^0 \rightarrow \pi^0\pi^0\gamma,
\phi \rightarrow K^+ K^-\gamma,\phi \rightarrow K^0\bar{K^0}\gamma$ are
predicted. The process $\phi \to K^0 \bar{K^0} \gamma$ is relevant to
precision measurment of CP-violation parameters in the kaon systerm at a
$\phi$-factory. We give a complete estimate of the branch ratio for
this decay process by including scalar resonance $f_0, a_0$ poles,
nonresonant smooth amplitude and an abnormal parity process with
$K^*$ pole which hasn't been considered before. We conclude that processes
with intermediate $K^*$ do not pose a potential background problem for 
$\phi\rightarrow K^0\bar{K}^0$ CP violation experiments.
\end{abstract}

\newpage
\section{Introduction}
Radiative decays $V \rightarrow P\bar P \gamma$(where V denotes vector
mesons, P denotes pseudoscalar mesons) have attracted much interest in
past decade[2--10]. The study of this kind of rare decays is important in
hadron physics, both because it is intimately related to the QCD-inspired
descriptions of the dynamics of mesons and because it has been urged
by experiments. For example, the reaction
$\phi\rightarrow K^0\bar{K^0}\gamma$ poses a possible backgroud problem of
$\phi\rightarrow K^0\bar{K^0}$ at future $\phi$ factory. The latter
process has been proposed as a way to study CP violation{\cite{b0}}.

The purpose of our present paper is to systematically study the processes
$\rho^0\rightarrow \pi^+\pi^-\gamma ,\pi^0\pi^0\gamma$,
$\phi \rightarrow K^+K^-\gamma ,K^0 \bar{K^0}\gamma$
in the framework of U(3)$_L\times$U(3)$_R$ chiral theory of 
mesons{\cite {Li1}}. In this effective chiral model, all couplings 
among pseudoscalars, and its lowest resonances are
fixed by introducing an universal coupling constant $g$. Thus, 
a unified description of mesons in the low energy is provided.

In fact, this effective model is an extended chiral quark model including 
$0^-, 1^\pm$ mesons. The chiral quark model,  originated by
Weinberg\cite{Wein79}, and then developed by Manohar and Georgi\cite{MG},
 provides a QCD-inspired description on the simple constituent quark
model. In the view of Manohar-Georgi model, between the chiral symmetry
breaking scale($\Lambda_{\chi SB}\sim 1-2$GeV) and the confinement scale
($\Lambda_{QCD} \sim 0.1-0.3$GeV), the dynamical field freedom are
constituent quarks(quasi-particle of quarks), gluons and Goldstone bosons
associated with chiral symmetry spontaneously breaking. In this
quasiparticle description, the effective gluon coupling is small and 
interactions between quarks and Goldstone bosons is important. 
The external gauge fields(e.g., photon field) can be introduced by
localizing the global chiral symmetry. On the other hand, it is well known
that in the electromagnetic interaction of mesons, the vector mesons play 
an essential role through VMD(Vector Meson Dominate)\cite{Sa}. Therefore, 
it is quite nature to extend chiral quark model to include spin-1 meson 
resonances via VMD and via minimal coupling principle.   
 
 The U(3)$_L\times$U(3)$_R$ chiral theory of mesons has been studied
extensively{\cite {Li1,Li2,G}}. The basic inputs for it are the 
pseudoscalar decay constants $f_P$, vector meson mass $m_V$ and a
universal coupling constant $g$. Predictions of this model are 
in good agreement with data{\cite {Li1,Li2,G}}. In particular, 
in Ref.{\cite{W}} it has been shown that the low energy limit of this
theory is equivalent to the chiral perturbation theory, and
the QCD constraints in Ref.{\cite {Ec}} are satisfied by this model.
Therefore, as an effective model of QCD, the U(3)$_L\times$U(3)$_R$
chiral theory of mesons is reliable.

The content of this paper is organized as follows.
In Sec.2, we present a brief review of chiral quark model and the basic
notations of the U(3)$_L\times$U(3)$_R$ chiral theory of mesons.
In Sec.3 and Sec.4, the branch ratio for these decays
are calculated respectively, and the gauge invariance of these decay
amplitudes is checked explicitly. We give a summary of the results in
Sec.5

\section{Chiral quark model and U(3)$_L\times$U(3)$_R$ chiral theory of
mesons}
The simplest parametrization of chiral quark model is\cite{MG} 
\begin{equation} 
{\cal L}_{\chi}=\bar{\psi}(x)(i\gamma\cdot\partial-mu(x))\psi(x) 
\end{equation}   
where 
\begin{eqnarray*} 
\psi&=&(u,d,s)^T, \\
u(x)&=&{1\over 2}(1-\gamma_5)U(x)+{1\over 2}(1+\gamma_5)U^{\dag}, \\ 
U(x)&=&\exp (2i\lambda_a \Phi_a/f_0), 
\end{eqnarray*}   
and $\lambda_a$ are Gell-Mann matrices of SU(3), $\Phi_a$ are fields 
of pseudoscalar meson octet and $f_0 \simeq f_\pi=186MeV$, $m$ 
is a parameter related to the quark condensate.

This Lagrangian is invariant under global chiral symmetry transformation.
 The vector(${\cal V}_\mu$) and axial-vector(${\cal A}_\mu$) external
field are introduced into ${\cal L}_\chi$ due to the requirment of
local chiral symmetry, i.e., we can replace the derivative operator
in Eq.(1) by covariant derivative operator with affine connection(or
gauge potential) ${\cal V}_\mu+{\cal A}_\mu \gamma_5$ as follows:
\begin{eqnarray}
\partial_\mu \rightarrow \tilde \an_\mu=\partial_\mu-i({\cal V}_\mu+{\cal
A}_\mu \gamma_5).
\end{eqnarray}
Then we obtain a theory describing strong and elecro-weak interactions
of mesons. 

Spin-1 meson resonances can be included via VMD, i.e., via substitution of
a new affine connection $(({\cal V}+v)_\mu+({\cal A}+a)_\mu \gamma_5)$ 
for the former one  ${\cal V}_\mu+{\cal A}_\mu \gamma_5$ in Eq.(2),
\begin{eqnarray}
\tilde\an_\mu \rightarrow \tilde{\tilde\an}_\mu\equiv \partial_\mu -
i(({\cal V}+v)_\mu+({\cal A}+a)_\mu \gamma_5),
\end{eqnarray}
with
\begin{eqnarray}
& &a_{\mu}=\tau_{i}a^{i}_{\mu}+\lambda_\alpha K^\alpha_{1\mu}+(\frac{2}{3}
+\frac{1}{\sqrt{3}}  
\lambda_8)f_{\mu}+(\frac{1}{3}-\frac{1}{\sqrt{3}}
\lambda_8)f_{s\mu},\nonumber\\
& &v_{\mu}=\tau_{i}\rho^{i}_{\mu}+\lambda_\alpha
K_{\mu}^{*\alpha}+(\frac{2}{3}+
\frac{1}{\sqrt{3}}\lambda_8)\omega_{\mu}+(\frac{1}{3}-
\frac{1}{\sqrt{3}}\lambda_8)\phi_{\mu},
\end{eqnarray}
where $i$=1, 2, 3 and $\alpha$=4, 5, 6, 7.

Now, the author of Ref.\cite{Li1} came to this extention and proposed a
Lagrangian,
\begin{eqnarray}
{\cal L}&=&\bar{\psi}(x)(i\gamma\cdot(\partial-i((
e_0Q A+v)+a\gamma_{5}))-mu(x))\psi(x) 
+{1\over 2}m_v^2 (v_{\mu a}v^\mu_a + a_{\mu a} a^\mu_a),
\end{eqnarray}
where $A_\mu$ is photon field, $Q$ is the
electric charge. The mass term of $v$ and 
$a$ is chiral gauge invariant because the $v$ and $a$ transform
homogeneously under local $U(3)_L\times U(3)_R$ symmetry. 
Note that there are no kinetic terms in Eq.(5) for all meson fields,
since they are treated as composited fields of quark fields instead of the
fundamental fields. The kinetic terms for these fields will be generated
via loop effects of quarks.

Following Ref. \cite{Li1}, the effective Lagrangian of mesons (indicated
by ${\cal M}$) are obtained through integrating over the quark fields,
\begin{equation}
{\rm exp}\{i\int d^4 x {\cal L^M}\}=
\int [d\psi][d\bar{\psi}]{\rm exp}\{i\int d^4 x {\cal L}\}.
\end{equation}
Using the dimensional regularization, and in the chiral limit,
 the effective Lagrangian ${\cal L}_{RE}$ (normal parity
part) and ${\cal L}_{IM}$ (abnormal parity part) have been evaluated
in Refs.\cite{Li1}. The Lagrangian describing normal parity processes
reads
\begin{eqnarray} \label{lag}
{\cal
L}_{RE}&=&{F^2\over 16}TrD_{\mu}UD^{\mu}U^\dagger
-{\frac{g^2}{16}}Tr(L_{\mu\nu}L^{\mu\nu}+R_{\mu\nu}R^{\mu\nu})\nonumber\\
& &+i\frac{N_C}{2(4\pi)^2}Tr(D_{\mu}UD_{\nu}U^{\dagger}L_{\nu\mu}+
D_{\mu}^{\dagger}UD_{\nu}UR_{\nu\mu})
+\frac{N_C}{6(4\pi)^2}TrD_{\mu}D_{\nu}UD^{\mu}D^{\nu}U^\dagger\nonumber\\
& &-\frac{N_C}{12(4\pi)^2}
   Tr(D_{\mu}UD^{\mu}U^{\dagger}D_{\nu}UD^{\nu}U^\dagger+
   D_{\mu}U^{\dagger}D^{\mu}UD_{\nu}U^{\dagger}D^{\nu }U-
   D_{\mu}UD^{\nu}U^{\dagger}D_{\mu}UD^{\nu}U^\dagger)\nonumber\\
& &+{1\over 8}m_0^2Tr(L_{\mu}L^{\mu}+R_{\mu}R^{\mu}),
\end{eqnarray}
where
\begin{eqnarray}
L_\mu&=&v_\mu-a_\mu \;,\;\;\;
R_\mu=v_\mu+a_\mu \;,\nonumber\\
D_{\mu} U&=&\an _\mu U-iL_\mu U+iUR_\mu,\nonumber\\
D_{\mu} U^\dagger&=&\an_\mu U^\dagger-iR_\mu U^\dag+iU^\dag L_\mu, 
\nonumber\\
L_{\mu\nu}&=&\an_{\mu}L_\nu-\an_{\nu}L_\mu-i[L_\mu,L_\nu]+e_0Q
F^L_{\mu\nu}\;,\nonumber\\
R_{\mu\nu}&=&\an_{\mu}R_\nu-\an_{\nu}R_\mu-i[R_\mu,R_\nu]+e_0Q
F^R_{\mu\nu}\;, \nonumber\\
\an_{\mu}\Psi&=&\partial_{\mu}\Psi+[-ie_0QA_{\mu},\Psi]
\;\;\;\;\; \Psi=U,U^\dagger,L,R,\; \nonumber\\
F^L_{\mu\nu}&=&F^R_{\mu\nu}=\partial_\mu A_\nu-\partial_\nu A_\mu\;.
\end{eqnarray}
Here an universal coupling constant $g$ has been introduced to absorb the
logarithmic divergence due to the integral of quark loop.

In Lagrangian (~\ref{lag}) the field $a_{\mu}(x)$ mixes with
$\partial_\mu{\Phi(x)}$, which 
should be diagonalized conveniently via field redefinition,
\begin{eqnarray}\label{sub}
 L_\mu \rightarrow L_\mu +i\frac{c}{g}U\an_\mu U^\dagger\;, \nonumber\\
 R_\mu \rightarrow R_\mu +i\frac{c}{g}U^\dagger \an_\mu U\;. 
\end{eqnarray}
Then the following equations are derived,
\begin{eqnarray}
& &\frac{F^2}{f_\pi^2}(1-\frac{2c}{g})=1,\;\;\;
c=\frac{f_\pi^2}{2gm_\rho^2},\;\;\;\mbox{for two-flavor}.\\
& &\frac{F^2}{f_K^2}(1-\frac{2c'}{g})=1,\;\;\;
c'=\frac{f_K^2}{2gm_{K^*}^2},\;\;\;\mbox{for three-flavor}
\end{eqnarray}

It should be stressed that this field redefinition is different from
the one in Ref.\cite{Li1}, $a_\mu \rightarrow
a_\mu-\frac{c}{g}\partial_\mu\Phi$. Eq.(~\ref{sub}) keeps the chiral
symmetry explicitly. It plays an important role when we
check the electromagnetic gauge invariance of the decay amplitude in
following sections. Eq.(7)-Eq.(11) provide the formalism employed in this
paper and all the calculations are performed in the chiral limit.

\section{The decay $\rho^0 \rightarrow \pi^+\pi^-\gamma,\pi^0\pi^0\gamma$}

In this section, we will restrict our calculations to the two-flavor case. 
The needed vertices to evaluate these processes can be obtained from
Sec.2. For the reaction $\rho^0 \rightarrow \pi^+\pi^-\gamma$, 
the Feynman diagrams are shown in Fig.\ref{figure:0(a)},
the involved vertices are:
\begin{figure}[h]
   \centering
   \psfig{figure=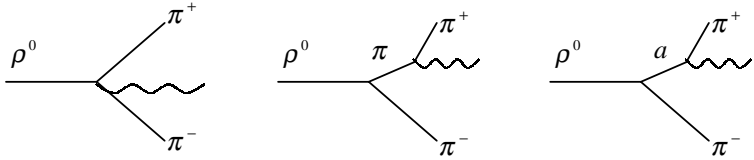,height=1.2in}
   \caption{$\rho^0 \rightarrow \pi^+ \pi^-\gamma$}
   \label{figure:0(a)}
\end{figure}
\begin{eqnarray}
{\cal L}_{\rho\pi\pi\gamma}
&=&eA_1^\rho\rho_{\mu}A^\mu\pi^i\pi^i+
eA_2^\rho\partial_\mu\rho_\nu F^{\mu\nu}\pi^i\pi^i{\;},\\
{\cal L}_{\rho\pi\pi}&=&
A_1^\rho
\rho_{\mu}^k\pi^i\partial^\mu\pi^j\epsilon_{ijk}{\;}, \\
{\cal L}_{\rho{a}\pi}&=&\gamma\epsilon_{ijk}[ 
B_1^\rho
\partial_\nu\rho_\mu^k a^{\mu j}\partial^\nu\pi^i +
B_2^\rho
\partial_\mu\rho_\nu^k a^{\mu i}\partial^\nu\pi^j \nonumber\\
& &+B_3^\rho
\rho_{\mu}^k a_{\nu}^i\partial^{\mu\nu}\pi^j
+B_4^\rho
\partial^2\rho_\mu^k a^{\mu i}\pi^j],\\
{\cal
L}_{\gamma\pi\pi}&=&eA_\mu\pi^i\partial^\mu\pi^j\epsilon_{ijk}{\;},\\
{\cal L}_{\gamma a \pi}&=&e\frac{\gamma f_\pi}{2g\pi^2
F^2}F_{\mu\nu}a^{\mu i}\partial^{\nu}\pi^j
\epsilon_{ijk}{\;},
\end{eqnarray}
where
\begin{eqnarray}
A_1^\rho&=&\frac{2}{g}
\{1+\frac{m^2_\rho}{2\pi^2 f^2_\pi}[(1-\frac{2c}{g})^2-4\pi^2
c^2)]\},\;\;\;
A_2^\rho=\frac{c-g} {g^2\pi^2 F^2}, \nonumber \\
B_1^\rho&=&-\frac{f_\pi}{g^2\pi^2 F^2}+\frac{8c}{g f_\pi \gamma^2},
\;\;\;\;\;\;\;
B_2^\rho=\frac{4c}{g f_\pi}+\frac{3f_\pi}{g^2\pi^2 F^2}, \nonumber \\
B_3^\rho&=&\frac{4c}{g f_\pi}+\frac{2f_\pi}{g^2\pi^2 F^2},
\;\;\;\;\;\;\;\;\;
B_4^\rho=\frac{1}{g^2\pi^2 f_\pi} ,\nonumber \\
\gamma&=&(1-\frac{1}{2g^2\pi^2})^{-\frac{1}{2}}.
\end{eqnarray}
The decay amplitude of this process is:
\begin{eqnarray}
{\cal M}_{\rho^0 \rightarrow \pi^+\pi^-\gamma}&=&<\pi^+\pi^-\gamma\vert
T\;\exp(i\int d^4 x ({\cal L}_{\rho\pi\pi\gamma}(x)+{\cal
L}_{\rho\pi\pi}(x) 
+{\cal L}_{\rho{a}\pi}(x)+{\cal L}_{\gamma\pi\pi}(x)+{\cal L}_{\gamma a
\pi}(x) )\vert \rho > \nonumber \\
&=&{\cal M}_a+{\cal M}_b+{\cal M}_c \;\;,
\end{eqnarray}
where
\begin{eqnarray}
{\cal M}_a&=&<\pi^+\pi^-\gamma\vert 
iT\;\int d^4 x {\cal L}_{\rho\pi\pi\gamma}(x)\vert \rho^0 > ,\\
{\cal M}_b&=&<\pi^+\pi^-\gamma\vert i^2T\;\int d^4 x\int d^4
y{\cal L}_{\rho\pi\pi}(x){\cal L}_{\gamma\pi\pi}(y)\vert \rho^0 > ,\\
{\cal M}_c&=&<\pi^+\pi^-\gamma\vert i^2T\;\int d^4 x\int d^4
y{\cal L}_{\rho a\pi}(x){\cal L}_{\gamma a\pi}(y)\vert \rho^0 >. 
\end{eqnarray}
Since photon is on-shell in this process, we should check the gauge
invariance of this amplitude. From Eq.(16) we see that 
the vertex of ${\cal L}_{\gamma a \pi}$ is already gauge invariant,
therefore, we need to check only the sum of ${\cal M}_a$ and ${\cal M}_b$. 
One can derive from Eqs.(19--20) that
\begin{eqnarray}
{\cal M}_a&=&2ie(A_1^\rho g_{\mu\nu}+A_2^\rho p\cdot q g_{\mu\nu}
-A_2^\rho q_\mu p_\nu) \varepsilon^{\mu}_{\vec p}e^{\nu}_{\vec q}\;,
\nonumber \\
{\cal M}_b&=&2ieA_1^\rho (\frac{k_{\mu}^+ k_{\nu}^-} {q \cdot k^-}+
\frac{k_{\mu}^- k_{\nu}^+} {q \cdot k^+})  \varepsilon^{\mu}_{\vec
p}e^{\nu}_{\vec q}\;, 
\end{eqnarray}   
where $p,k^+,k^-,q$
denote the momenta of $\rho^0,\pi^+,\pi^-$ and photon fields respectively, 
and $\varepsilon^{\mu}_{\vec p}, e^{\nu}_{\vec q}$ are the polarization
vectors for $\rho$ meson and photon field respectively.
If we substitute $e^{\nu}_{\vec q}$ with $q^\nu$ in Eq.(22), we will
obtain:
\begin{eqnarray}
({\cal M}_a+{\cal M}_b){\vert}_{e^{\nu}_{\vec q} \rightarrow q^\nu}\propto
(q_\mu+k^+_\mu+k^-_\mu)\varepsilon^{\mu}_{\vec p}{\;}.
\end{eqnarray}  
Using four-momentum conservation and the space-like condition of
the wave function of vector field, 
\begin{eqnarray}
& &p=q+k^+ + k^- {\;},\nonumber\\
& &p\cdot \varepsilon_{\vec p}=0{\;}.
\end{eqnarray}
Eq.(22) vanishes, so the gauge invariance of this amplitude is kept.

\begin{figure}[]
\centering
\psfig{figure=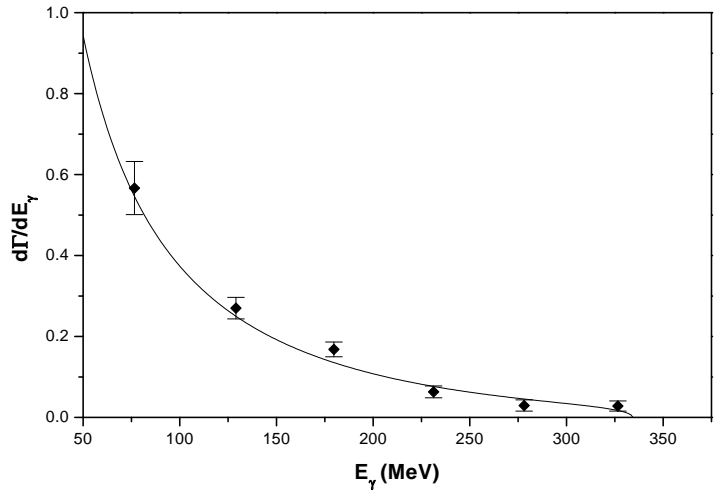,height=3in}
\caption{Photon spectrum,$d\Gamma/dE_\gamma$,for the process $\rho^0
\to \pi^+ \pi^- \gamma$. The experimental data taken from Ref.[20]
 are normalized to our results} 
\label{figure:1(a)}
\end{figure}

Before we give the numerical results of the width and the branch ratio of
this process, it is necessary to point out that there is 
no adjustable parameter in our calculation. The basic input $g$, as an
universal coupling constant in this theory, can be fixed by a experiment, 
thus we can compare our theoretical results with experimental data and 
give predictions for processes which have not
been measured in experiments.
In this paper, we take $g=0.39$. Then we obtain,
\begin{eqnarray*}
& &\Gamma(\rho^0 \rightarrow \pi^+\pi^-\gamma)=1.54\;\; \mbox{MeV}{\;}, \\
& &B(\rho^0 \rightarrow \pi^+\pi^-\gamma)=1.03\times10^{-2}
\;\;\;\;\;\;\;\mbox{for}\;\;E_\gamma >50\;\;\mbox{MeV}.
\end{eqnarray*}  
which compares favourably with the experimental data{\cite {Do}}, $B^{exp} 
(\rho^0 \rightarrow \pi^+\pi^-\gamma)=(0.99\pm 0.04\pm 0.15) 10^{-2}$
for $E_\gamma >50$MeV, where $E_\gamma$ is the photon energy in the rest
frame of $\rho^0$. The shape of differential spectrum with respect to
photon energe is also given in the Fig. \ref{figure:1(a)}. One can see
that the experimental result is in good agreement with our theoretical
expectations.

\begin{figure}[h]
   \centering
   \psfig{figure=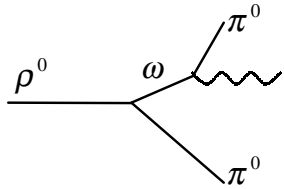,height=1.2in}
   \caption{$\rho^0 \rightarrow \pi^0 \pi^0\gamma$}                     
   \label{figure:0(b)}
\end{figure}

\begin{figure}[]
   \centering
   \psfig{figure=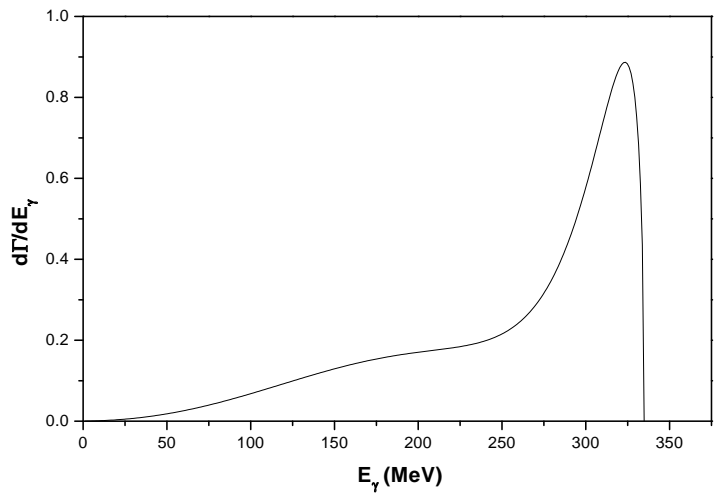,height=3in}
   \caption{Photon spectrum, $d\Gamma/dE_\gamma$, for the process $\rho^0
            \rightarrow \pi^0\pi^0\gamma$.}
   \label{figure:1(b)}
\end{figure}

The reaction $\rho^0 \rightarrow \pi^0\pi^0\gamma$
involves only abnormal parity, the Feynman diagrams is shown in 
Fig.\ref{figure:0(b)}. Following Ref.{\cite {Li1}}, by means of
bosonization the quark propagator, we obtain               
\begin{eqnarray}
& &{\cal L}_{\rho\omega\pi}=-\frac{3}{\pi^2 g^2 f_\pi}\varepsilon^
{\mu\nu\alpha\beta}\partial_\mu\omega_\nu\rho_\alpha^i\partial_\beta
\pi^i {\;},\\
& &{\cal L}_{\gamma\omega\pi}=-e \frac{3}{2 \pi^2 g f_\pi}\varepsilon^
{\mu\nu\alpha\beta}\partial_\mu\omega_\nu A_\alpha\partial_\beta\pi^0{\;}.
\end{eqnarray}
The vertex ${\cal L}_{\gamma\omega\pi}$ is obviously gauge invariant
because of the totally antisymmetry tensor
$\varepsilon^{\mu\nu\alpha\beta}$. The branch ratio is calculated to
be 
\begin{eqnarray}
B(\rho^0 \rightarrow \pi^0\pi^0\gamma)=1.02\times 10^{-5} {\;},
\end{eqnarray}
and the differential spectrum with respect to photon energy is shown 
in Fig.\ref{figure:1(b)} which will be tested in future experiments.

\section{The decay $\phi\rightarrow K \bar{K}\gamma$}
\begin{figure}[h]
   \centering
   \psfig{figure=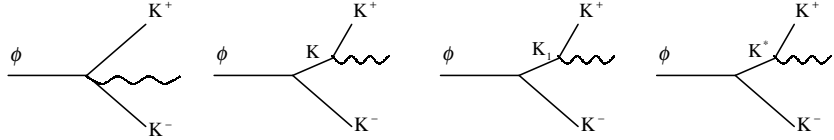,height=1.2in}
   \caption{$\phi \rightarrow K^+ K^-\gamma$}                     
   \label{figure:0(c)}
\end{figure}

In order to calculate the $\phi$ decay, we need to extend our calculation
to three-flavor case. First, we derive the vertices for the transition
$\phi\rightarrow K^+ K^- \gamma$, the Feynman digrams are shown in
Fig.\ref{figure:0(c)}, the vertices of this  process have both normal
parity part and abnormal parity part. The normal parity part is
\begin{eqnarray}                       
{\cal L}_{\phi K^+ K^- \gamma}
&=&eA_1^\phi\phi_\mu A^\mu K^+K^- 
{\;},\\
{\cal L}_{\phi K^+ K^-}&=&i A_1^\phi\phi_\mu K^+\partial^\mu K^- 
{\;},\\
{\cal L}_{\gamma K^+K^-}&=&ieA_\mu K^+\partial^\mu K^- {\;},\\
{\cal L}_{\phi K_1 K}&=&\gamma[
iB_1^\phi\partial_\nu\phi^\mu K_{1\mu}^- \partial^\nu K^+ +
iB_2^\phi\partial_\mu\phi^\nu K_{1\mu}^- \partial^\nu K^+ + \nonumber \\
& &iB_3^\phi\phi_\mu K_{1\nu}^-\partial^{\mu\nu}K^+ +
iB_4^\phi\partial^2\phi_\mu K_1^{\mu -}K^+]+ h.c. {\;\;},\\
{\cal L}_{\gamma K_1
K}&=&ie\frac{\gamma f_K}{2g\pi^2 F^2}F_{\mu\nu}K_1^{+\mu}\partial^\nu K^-
+ h.c.{\;\;},
\end{eqnarray} 
where
\begin{eqnarray}
A_1^\phi&=&-\frac{\sqrt2 m_\phi^2}{f_K^2}
[4c'(1-{c'\over g})+{1\over {g\pi^2}}(1-{2c' \over g})^2]
,{\;\;\;\;}
\nonumber \\
B_l^\phi&=&\frac{\sqrt 2}{2}{B_l^\rho (c\rightarrow c',f_\pi \rightarrow
f_K)}, {\;\;} l=1,2,3,4.
\end{eqnarray}
and the abnormal parity part is
\begin{eqnarray}
{\cal L}_{\phi K^* K^{\pm}}&=&
-\frac{3 \sqrt 2}{2\pi^2 g^2 f_K}\varepsilon^{\mu\nu\alpha\beta}
\partial_\nu\phi_\alpha K^{*+}_\mu\partial_\beta K^- +h.c.{\;\;},\\
{\cal L}_{\gamma K^* K^{\pm}}&=&
- \frac{e}{2\pi^2 g f_K}\varepsilon^{\mu\nu\alpha\beta}
\partial_\nu A_\alpha K^{*+}_\mu\partial_\beta K^-  + h.c.{\;\;},
\end{eqnarray}

\begin{figure}[]
   \centering
   \psfig{figure=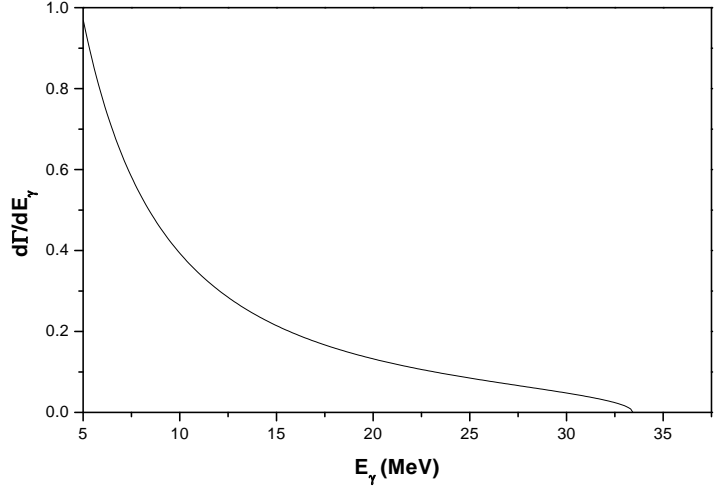,height=3in}
   \caption{Photon spectrum, $d\Gamma/dE_\gamma$, for the process $\phi
            \rightarrow K^+ K^-\gamma$. For $E_\gamma>5$MeV}
   \label{figure:1(c)}
\end{figure}

The gauge invariance can also be checked as in the $\rho$ decay. Using the
same value of $g$, i.e., $g=0.39$, we give the differential spectrum
in Fig.\ref{figure:1(c)} .
   
Now, we consider the transition $\phi \rightarrow K^0\bar{K^0}\gamma$. The
photon in this reaction is very soft($E_{max}<25$Mev), so it is difficult
to
distinguish it from the genuine $\phi \rightarrow K^0\bar{K^0}$ events
which has been proposed as a way to measure the small parameter in
studying CP violation. The branch ratio of $\phi \rightarrow
K^0\bar{K^0}\gamma(\stackrel{>}{\sim} 10^{-6})$ will limit the presion of
this
measurment. This quantity has been predicted by several authors[4-10]. 
In Refs.[4--8], the contribution of interchanging the scalar meson
$S(a_0,$or $f_0$) has been obtained via
chain reaction $\phi \rightarrow S\gamma \rightarrow K^0\bar{K^0}\gamma$,
in
which the decay $\phi \rightarrow S\gamma$ is proceeds through the charged
$K$ loop. The uncertainty of this approach is that the coupling constant
$g_{SK\bar K}$ is not well know because of the lack of experiment data. 
In Ref.\cite{b4}, the non-resonant contribution has been calculated using
current algebra and low energy theorem. In Ref.\cite{b6}, the authors
did not
introduce $a_0,f_0$ explicitly, they calculated the final state
intereaction of $KK$ system in a chiral unitary approach. This approach
generates the $a_0,f_0$ meson dynamically, the obtained amplitude after
summing over an infinite series diagram also contain non-resonant
contribution. All these calculations, however, did not contain the
contribution of an abnormal parity
process via interchanging $K^*$ which in principle must be added to the 
resonant poles and the nonresonant smooth amplitude, otherwise one may
not assume scalar meson dominance a priori. In the present paper, we
provide a complete calculation on the branch ratio for this decay by
including the abnormal parity process with $K^*$ pole, scalar resonance
$f_0, a_0$ poles and non-resonant amplitude. The role of scalar resonance
will be dealt with as in the former works[4--8,10]. The difference between
their scheme and ours is that the needed vertices to calculate the loop
diagram $\phi \to K^+K^-\gamma$ has been derived in Eqs.(28--30) as well 
as all the coupling constants has been fixed by the universal constant
$g$, in other words, there is no adjustable parameter in our calculation.  
The complete intereaction of this process including normal and
abnormal parity vertices is:
\begin{eqnarray}
{\cal L}&=&eA_1^\phi\phi_\mu A^\mu K^+K^-
+i A_1^\phi\phi_\mu K^+\partial^\mu K^-
+ieA_\mu K^+\partial^\mu K^- +{\cal L}_{\phi K^* K^0}+
{\cal L}_{\gamma K^* K^0}
\end{eqnarray}
where
\begin{eqnarray}
{\cal L}_{\phi K^* K^0}&=&
-\frac{3 \sqrt 2}{2\pi^2 g^2 f_K}\varepsilon^{\mu\nu\alpha\beta}
\partial_\nu\phi_\alpha K^{*0}_\mu\partial_\beta\bar{K^0} + h.c. {\;\;},\\
{\cal L}_{\gamma K^* K^0}&=&
\frac{e}{\pi^2 g f_K}\varepsilon^{\mu\nu\alpha\beta}
\partial_\nu A_\alpha K^{*0}_\mu\partial_\beta\bar{K^0} + h.c.{\;\;},
\end{eqnarray}
described the abnormal parity process. Denoting the momenta of $K^0,\bar
{K^0}$ as $k_1, k_2$ and defining $s_1=(k_1+q)^2, s_2=(k_2+q)^2$, we
derived the amplitude for this abnormal diagram(shown in
Fig.\ref{figure:2(a)}):
\begin{figure}[hpt]
   \centering
   \psfig{figure=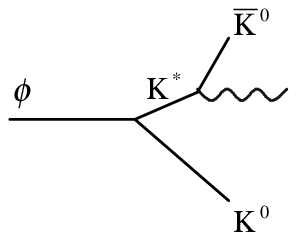,height=1.2in}
   \caption{$\phi \to K^0 \bar {K^0} \gamma$ through $K^*$.
}
   \label{figure:2(a)}
\end{figure}
\begin{eqnarray}
{\cal M}_1&=&<K^0 \bar{K^0} \gamma \vert i^2 T\;\int d^4 x \int d^4 y
{\cal L}_{\phi K^* K^0}(x){\cal L}_{\gamma K^* \bar{K^0}}(y) \vert
\phi>,\nonumber \\
&=&ie\frac{3\sqrt 2}{2\pi^4 g^3 f_K^2}
(d_0 g_{\mu \nu}+d_1 k_{1\mu} k_{1\nu} 
                                + \tilde {d_1} k_{2\mu} k_{2\nu}
                                +d_2 k_{1\mu} k_{2\nu}
                                + \tilde {d_2} k_{2\mu} k_{1\nu})
{\;}\epsilon^{\mu}_{\vec p}e^{\nu}_{\vec q} 
\end{eqnarray}
with
\begin{eqnarray}
d_0&=&\frac{s_1}{4(s_1-m_{K^*}^2)}(4k_1 \cdot k_2+s_1-m_\phi^2)
        +{\;} (k_1 \leftrightarrow k_2) ,\nonumber \\
d_1&=&{1 \over 2}(\frac{s_2}{s_2-m_{K^*}^2}
                 -\frac{m_\phi^2-s_1}{s_1-m_{K^*}^2}),{\;\;}
\tilde {d_1}=d_1(k_1 \leftrightarrow k_2),\nonumber \\
d_2&=&\frac{s_2}{s_2-m_{K^*}^2}-\frac{k_1 \cdot k_2}{s_2-m_{K^*}^2},
{\;\;}\tilde {d_2}=d_2(k_1 \leftrightarrow k_2).
\end{eqnarray}

In order to derive the contribution of scalar resonance $f_0$ and $a_0$,
we need to calculate the one loop diagram though $K^+K^-$ and the final
state intereactions of $K^+K^-$ to $K^0 \bar{K^0}$ shown in
Fig.\ref{figure:2} .Similar to Ref.\cite{b6}, we get the amplitude:
\begin{figure}[hpt]
   \centering
   \psfig{figure=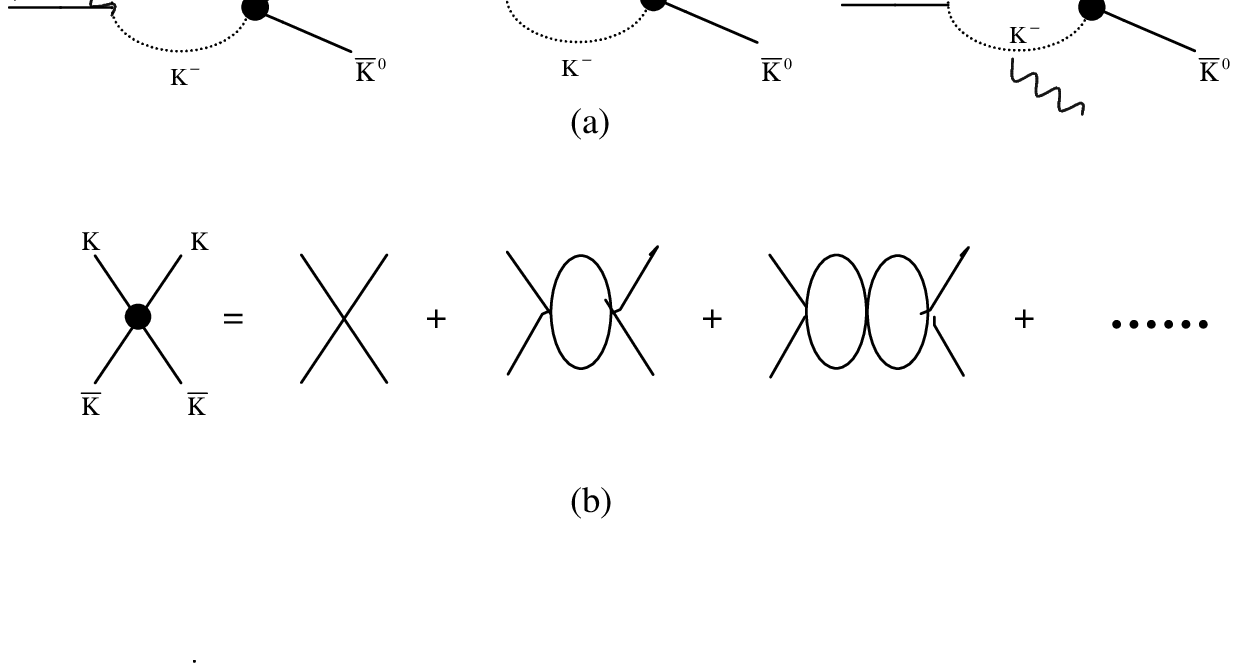,height=2.4in}
   \caption{(a).$\phi \to K^0 \bar {K^0} \gamma$ through charged $K$ loop.
(b).$K^+K^- \to K^0 \bar{K^0}$ amplitude, the intermediate
loops contain $\pi \pi, K\bar K, \pi^0 \eta$. 
}
   \label{figure:2}
\end{figure}
\begin{eqnarray}
{\cal M}_2&=&\frac{eA_1^\phi}{2\pi^2 i m_K^2}I(a,b)
[p\cdot q g_{\mu\nu}-p_\nu q_\mu]{\;}
\epsilon^{\mu}_{\vec p}e^{\nu}_{\vec q} {\;}
t_s
\end{eqnarray}
where
$a={m_\phi^2}/{m_K^2},b={Q^2}/{m_K^2},Q^2=(k_1+k_2)^2$,
\begin{eqnarray}
I(a,b)&=&\frac{1}{2(a-b)}-\frac{2}{(a-b)^2}(f({1\over b})-f({1\over a}))
                         +\frac{a}{(a-b)^2}(g({1\over b})-g({1\over a}))
\end{eqnarray}
with
\[
f(x)=\left\{
\begin{array}{l}
-\arcsin(\frac{1}{2\sqrt{x}})^2 {\;\;\;} x>{1\over 4} \\
{1\over 4}[\ln(\frac{\eta_+}{\eta_-})-i\pi ]^2 {\;\;\;} x<{1\over 4}
\end{array}
\right.
\]
\[   
g(x)=\left\{
\begin{array}{l}
(4x-1)^{1\over 2}\arcsin(\frac{1}{2\sqrt{x}})^2 {\;\;\;} x>{1\over 4} \\
{1\over 2}(1-4x)^{1\over 2}[\ln(\frac{\eta_+}{\eta_-})-i\pi ]^2 {\;\;\;}
x<{1\over 4}
\end{array}
\right.
\]
\begin{equation}
\eta_\pm={1\over{2x}}(1\pm \sqrt{1-4x})
\end{equation}
the $t_s$(Eq.(9) in \cite{b6}) is the scattering amplitude of $K^+K^-$ to
$K^0 \bar {K^0}$, its diagrammatic meaning is shown in
Fig.\ref{figure:2}(b), and its explicit expression can be obtained from
Eq.(30) in Ref.\cite{b7} which is derived from Lippmann-Schwinger equation
in the coupled channel approach. As declared in Ref.\cite{b6}, this
amplitude(Eq.(41)) also take into account nonresonant contribution. 

\begin{figure}[]
   \centering
   \psfig{figure=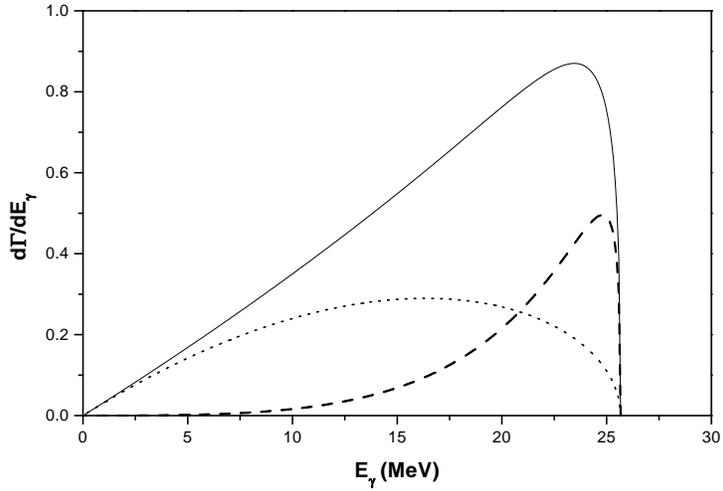,height=3in}
   \caption{Photon spectrum, $d\Gamma/dE_\gamma$, for the process
$\phi \rightarrow K^0\bar{K^0}\gamma$. Dot line: distribution only taking
into account contribution of abnormal parity process with $K^*$ poles,
dash line:distribution only taking into account the contribution of scalar
resonance poles and nonresonant amplitude, solid line: the total
distribution.}
   \label{figure:1(d)}
\end{figure}

The width and the branch ratio are given by: 
\begin{eqnarray}
\Gamma(\phi \rightarrow K^0 \bar{K^0} \gamma)&=&\frac{\alpha}{192\pi^2
m_\phi^3} \int ds_1 dQ^2 ({\cal M}_1+{\cal M}_2)^2 \\
B(\phi \rightarrow K^0 \bar{K^0} \gamma)&=&\Gamma(\phi \rightarrow K^0
\bar{K^0} \gamma)/4.43
\end{eqnarray}
where $\alpha=e^2/4\pi=1/137$. In the following numerical evaluation,
the constant $g$ take the same value 0.39 as in the previous. If we
neglect the abnormal parity
process, i.e. set ${\cal M}_1=0$, then we obtained $B(\phi \rightarrow 
K^0 \bar{K^0} \gamma)_{scalar}=5.6\times 10^{-8}$, which is a little
different with the value $5\times 10^{-8}$ 
quoted in Ref.\cite{b6}. On the other hand, if we neglect the contribution
of scalar  resonance, i.e. set ${\cal M}_2=0$, we obtained $B(\phi
\rightarrow K^0 \bar{K^0} \gamma)_{abnormal}=7.6\times 10^{-8}$. We see
that the contribution of this abnormal parity process is the same
important as the
scalar resonance poles, so its contribution can not be neglected.
After performing the integral of Eq.(44), we obtained:
\begin{eqnarray}
B(\phi \rightarrow
K^0\bar{K^0}\gamma)&=&1.8\times10^{-7}{\;}.
\end{eqnarray} 
We see although the interference is constructive, it will not provide much
significant background for precision test of CP-violation in $\phi
\rightarrow K \bar K$. The differential spectrum with respect to photon
energy for these three case(abnormal, scalar, interference) are given in 
Fig.\ref{figure:1(d)}.

\section{Summary}
To conclude, in this paper, we perform a systematic calculation of 
$\rho$ and $\phi$'s radiative decays in an extended chiral quark model,
 in which all coupling are fixed by the universal coupling constant 
$g$. The gauge invariance of these decay amplitudes has been checked. The
theoretical differential spectrum with respect to photon energy and
branch ratio of $\rho^0 \rightarrow \pi^+\pi^- \gamma$ agree
with the experimental data well. Predictions of differential spectrum
and branch ratio for the processes
$\rho^0 \rightarrow \pi^0\pi^0 \gamma, \phi \rightarrow K^+ K^- \gamma,
K^0 \bar {K^0} \gamma$ have been derived. The branch ratio for $\phi
\rightarrow K^0 \bar{K^0} \gamma$ including the contribution of abnormal
parity process with $K^*$ pole, scalar resonance $a_0, f_0$ poles and
nonresonant amplitude has been calculated to be about $10^{-7}$ and will
not limit the precision measurment of the small CP-violation parameters at
future $\phi$ factory.

\begin{center}
{\bf ACKNOWLEDGMENTS}
\end{center}
We would like to thank Dr. DaoNeng Gao for his helpful discussion.
This work is partially supported by NSF of China through C. N. Yang
and the Grant LWTZ-1298 of Chinese Academy of Science.

\newpage

\end{document}